\documentclass[aps,axodraw]{revtex4}
\usepackage{fancyhdr}
\usepackage{fancybox}
\usepackage{amsmath,amssymb,graphicx}

\newcommand{\beq}{\begin{eqnarray}}
\newcommand{\eeq}{\end{eqnarray}}

\newcommand{\drawsquare}[2]{\hbox{%
\rule{#2pt}{#1pt}\hskip-#2pt
\rule{#1pt}{#2pt}\hskip-#1pt
\rule[#1pt]{#1pt}{#2pt}}\rule[#1pt]{#2pt}{#2pt}\hskip-#2pt
\rule{#2pt}{#1pt}}

\newcommand{\fund}{\drawsquare{6.5}{0.4}}

\newcommand{\asymm}{\raisebox{-3.5pt}{\drawsquare{6.5}{0.4}\hskip-6.9pt%
        \raisebox{6.5pt}{\drawsquare{6.5}{0.4}}}}

\newcommand{\ov}{\overline}

\begin{document}


\title{ D-term Enhancement in Spin-1 Top Partner Model} 


\author{ Haiying Cai}


\affiliation{ Department of Physics, Peking University,  China.}


\begin{abstract}
Supersymmetric models with extended electroweak gauge groups have
the potential to enhance the Higgs quartic interaction through
nondecoupling D-terms. We consider the D-term enhancement effect
in a vector top partner model, where the quadratic divergence to
the Higgs mass from the virtual top quark is canceled by its
corresponding spin-1 superpartners. We are going to show that the
model can predict a Higgs mass beyond the LEP  bound, and is
consistent with the precision electroweak constraints.

\end{abstract}

\maketitle

\section{Introduction}

In the Supersymmetric theory, since the quadratic divergence
associated with the Higgs mass-squared from the SM fields will be
canceled by their superpartners,  soft SUSY breaking terms only
induce logarithmic corrections and the scalar field mass is
stabilized to be around the soft SUSY breaking scale $m_s$. In
order to let the SUSY theory to be natural and therefore reduce
fine-tunings, the soft SUSY breaking scale $m_s$ is supposed to be
in the hundred GeV range. In the minimal supersymmetric standard
model (MSSM),  the physical Higgs mass is related to the mass of
$Z$ gauge bosons times  a  factor of $\cos (2\beta) $ at the
tree level, where $\beta$ is determined by the ratio of the two
Higgs fields vacuum expectation values (VEVs) ($v_u/v_d $).  However,
the LEP direct search excluded the existence of  a Higgs bosons
below $114.4 ~\mbox{GeV}$ at $95\%$ C.L. For the Higgs to go beyond
the LEP bound, large radiative contribution to the quartic
interaction term from the top quark sector is necessary, which in
turn demands
 the top squark to  have a mass of the TeV order.  The tension between the electroweak
scale and the new physics emerging scale, which is referred as the
little hierarchy problem, encourages people to explore new
possibilities to avoid the dilemma. There are many attempts to
achieve a Higgs mass much heavier than the $Z$  gauge bosons in
the supersymmetric theory. One straightforward way is to enhance
the quartic interaction term at the tree level, and generally
additional interaction structure is required. In the NMSSM model
\cite{NMSSM} , one extra $SU(2)_L$ singlet superfield $N$ is
added, which couples with the two Higgs fields through a
supersymmetric Yukawa interaction $ \lambda N H_u H_d $. A large
$\lambda $ is preferred to generate a large quartic term but the
requirement that the Yukawa interaction is perturbative till the
unification scale  puts an upper bound on the Higgs mass. An
alternative method to raise the Higgs mass without inducing  fine
tunings is to consider a  fat Higgs scenario  originated from a
strong interaction sector . In the fat higgs scenario, the singlet
chiral field $N$ and the two Higgs fields $H_u$ and $H_d$ are
composite meson fields interacting  via a naturally large Yukawa
coupling.  The original fat Higgs model  has a dynamically
generated  superpotential $\lambda N (H_u H_d - v^2 ) $ with the
similar matter content as  the NMSSM  in the low energy scale
\cite{Fat}. This type of theory is further extended by Refs.
\cite{Chang} and \cite{Delgado}.   In the  New Fat Higgs  model
\cite{Chang},  only  the singlet chiral field $N$ is composite
while the two Higgs fields are still kept elementary.

Supersymmetric models with enlarged gauge groups under which the Higgs bosons are charged may raise the Higgs mass through the nondecoupling D-term effects \cite{Tait} \cite{Wacker}.  In the low energy scale, the enlarged gauge groups need to be broken into
the Standard Model gauge group by the VEVs of some extra scalar fields.  If the gauge symmetry is broken in a  SUSY conserving limit,  the D-term effects of these extra scalar fields would decouple and we could recover the standard MSSM D-term potential for the Higgs fields. In order to retain the D-term effects from those extra fields till the electroweak scale,  SUSY breaking  effects need to be included in the mechanism responsible for the gauge symmetry breaking.  When the SUSY breaking scale is much larger than the gauge symmetry breaking scale, the effective D-term for the Higgs fields in the electroweak scale can be enhanced.

In this Letter I consider the possibility to increase the Higgs
quartic interaction terms in a spin-1  top partner model
\cite{Cai}. In this model, the superpartners of the left-handed
top quark are spin-1 vector bosons. while the superpartner of the
right-handed top quark is still a  scalar. This scenario is
realized by extending the gauge group and assembling the
left-handed top quark into a vector supermultiplet. The extended
gauge group serves to provide the source of  nondecoupled D-term
effects. Extra chiral fields are necessary to be added to trigger
the gauge symmetry breaking since we hope to achieve a  D-term
flat minimum.   In the following of this letter,  I will specify
the  superpotential  responsible for  gauge symmetry breaking and
supersymmetry breaking. The exact mass spectrum for scalar states
in the link fields after the symmetry breaking will be calculated. 
I am going to verify the D-term enhancement effects in the Higgs
bosons sector and explore the  bound for the mass of  the Higgs
boson in this model after considering relevant electroweak
constraints.

\section{D-term Enhancement and Higgs Mass}

\begin{figure}[htb]
\begin{center}
\includegraphics[width=0.35\hsize]{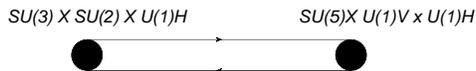}
\end{center}
\caption{two sites illustration for spin-1 top partner model \label{model}}
\end{figure}

We first briefly review the structure of the spin-1 top partner model. For details of the realization, one can refer to the previous paper \cite{Cai}. The model is based on the gauge group $ SU(3) \times SU(2) \times U(1)_H  \times U(1)_V \times SU(5) $, which can be better illustrated in a supersymmetric two sites moose diagram (See Fig~[\ref{model}]).  One copy of  three generations of leptons, quarks plus their superpartners
are put in the first moose site with a gauge group of $ SU(3) \times SU(2) \times U(1)_H$.  These chiral superfields  transform exactly  the same as in the MSSM. While two higgs superfields $H$ and $\bar H$ need to be put in a second moose site which has a $SU(5) \times U(1)_H \times U(1)_V$ gauge group.  The gauge coupling of $U(1)_{H} $ can be set to be very small.
Four vector-like link fields $\Phi_3$, $\ov{\Phi}_3$,
$\Phi_2$, $\ov{\Phi}_2$ are responsible to communicate between the fields located in the two isolated moose sites.
When the link fields gain nonzero VEVs, they break the original product gauge group
$SU(3) \times SU(2)\times U(1)_H \times SU(5)\times U(1)_V$ down to the
diagonal MSSM gauge group $ SU(3)_C \times  SU(2)_W \times U(1)_Y $.
The gauge transformation property of the Higgs fields and  the four link fields are given in Table~\ref{tab:content}.
%
\begin{table}
\begin{center}
\begin{tabular}{c|ccr|rc|c}
 & $SU(3)$ & $SU(2)$  & $U(1)_H$ & $U(1)_V$ & $SU(5)$ & $H+V+a T_{24}$\\
\hline
$H$ & ${\bf 1}$ & ${\bf 1}$ & $\frac{1}{2}$ & $\frac{1}{10}$\hphantom{0}   &\fund
&$(\frac{2}{3},\frac{1}{2})$
$\vphantom{\raisebox{3pt}{\asymm}}$
\\
${\overline H}$ & ${\bf 1}$ &  ${\bf 1}$ &  $-\frac{1}{2}$  & $-\frac{1}{10}$\hphantom{0}
&${\overline \fund}$ &$(-\frac{2}{3},-\frac{1}{2})$
$\vphantom{\raisebox{3pt}{\asymm}}$\\

$\Phi_3$ & ${\overline \fund} $& ${\bf 1}$  & $-\frac{1}{6}$ & $\frac{1}{10}$\hphantom{0}    &
\fund  &$(0,-\frac{1}{6})$
$\vphantom{\raisebox{3pt}{\asymm}}$\\
$\Phi_2$ & ${\bf 1}$ & ${\overline \fund}$ & $0$ & $\frac{1}{10}$\hphantom{0}
&\fund &$(\frac{1}{6},0)$
$\vphantom{\raisebox{3pt}{\asymm}}$\\

${\overline \Phi_3}$ &  \fund   & ${\bf 1}$ & $\frac{1}{6}$ & -$\frac{1}{10}$\hphantom{0}
&${\overline \fund}$ &$(0,\frac{1}{6})$
$\vphantom{\raisebox{3pt}{\asymm}}$\\

${\overline \Phi_2}$ & ${\bf 1}$ & \fund & $0$  & $-\frac{1}{10}$\hphantom{0}
&${\overline \fund}$&$(-\frac{1}{6},0)$
$\vphantom{\raisebox{3pt}{\asymm}}$\\

\end{tabular}
\end{center}
\caption{\label{tab:content}  The gauge structure of the two Higgs fields and the four link fields .}
\end{table}

The chiral superfields $\overline H=(\ov{T}, H_1) $ and $H=(\overline{T}^c, H_2)$ transform as fundamental and antifundamental under the $SU(5)$ gauge group respectively. $H_1 $ and $H_2 $ are the traditional up-type and down-type Higgs superfields in the MSSM.  Since the triplet companies of $H_1 $ and $H_2$ transform as $ (\bar 3, 1)$ and  $(3, 1 )$ under the diagonal $ SU(3)_C \times SU(2)_W $  subgroup, they are treated as one couple of vector-like right-handed heavy top quarks in our model.  The link fields can mix $\ov{T} $ with the right-handed field $\bar u_3 $ living in the first moose site,  so that one linear combination still has zero mass and thus is the physical right-handed top quark.  Both of the $U(1)_H $ and $U(1)_V $ charges are  necessary for arranging a correct hypercharge and a correct baryon number  for the Higgs bosons and top quarks.

For the vector supermultiplet in the $SU(5)$ gauge group, the
extra gauge bosons $X$, $Y$ transforming as $(3, 2) $ under the $
SU(3)_C \times SU(2)_W $ are identified as the spin-1 top partner
in this model. They are lifted to be heavy after the gauge
symmetry breaking  in the same way as in the GUT model .  We also
need to identify the field contents in the four link fields. Under
the diagonal $SU(3)_C \times SU(2)_W$, $\Phi_3$ ($\ov{\Phi}_3$)
will split into one complex singlet $\Phi_{3 S} $ ($\ov{\Phi}_{3
S} $), one complex octet  $\Phi_{3 O } $ ($\ov{\Phi}_{3 O} $),
plus one component field  $\Phi_{3 t}$ ($\ov{\Phi}_{3 t}$) with
the same quantum number as $\bar t_L $ ($ t_L$ ) in the first
moose site. Similar decomposition applies to the fields of
$\Phi_2$  and $\ov{\Phi}_2$.  They  will split into  one complex
singlet $\Phi_{2 S } $ ($\ov{\Phi}_{2 S} $), one complex triplet
$\Phi_{2 T } $ ($\ov{\Phi}_{2 T} $),  as well as one component
field $\Phi_{2t } $ ( $\ov{\Phi}_{2 t} $)  which has the same
quantum number as $t_L $  ( $ \bar t_L$ ).   $\Phi_{2 t } $ and
$\ov{\Phi}_{3 t} $ mix with the $(3, 2)$ sector gaugino
$\lambda_{32}$ through gauge interaction and they mix with the
left handed field $Q_3 = (t_L, b_L )$ in the first moose site
through Yukawa interaction.  As long as the $\mu$-terms for $
\Phi_{2,3} $ and $\ov{ \Phi}_{2,3}$ are large enough compared with
their VEVs,  the dominant component of the physical left handed
top quark will be the gaugino $\lambda_{32}$ and their
superpartner are spin-1 gauge bosons.

We now write down the superpotential relevant for calculations.
In order to get  quartic terms for the link fields, we may add two
singlets chiral superfields $S_{1,2}$ to interact with
$\Phi_{2,3}$ and $\ov{\Phi}_{2,3}$. One adjoint chiral superfield
$A_1$ charged under the $SU(2)$ gauge group and another adjoint
chiral superfield $A_2$ charged under the $SU(3)$ gauge group are
also added to ensure that there are no light modes after the gauge
symmetry breaking.
\beq
W_{susy} &=& y_1 Q_3 \Phi_3 \ov{\Phi}_2 + \mu_2 \Phi_2 \ov{\Phi}_2  + \mu_3 \Phi_3 \ov{\Phi}_3
+ \lambda_S S_1 ( \Phi_2 \ov{\Phi}_2 - w_2^2 ) + \lambda_S  S_2  ( \Phi_3 \ov{\Phi}_3-w_3^2)  \nonumber \\
& &+  \lambda_{A} \ov{\Phi}_2 A_1^a \frac{\sigma^a}{2} \Phi_2 + \lambda_{A} \ov{\Phi}_3 A_2^m G^m \Phi_3 .
\eeq
$\sigma^a /2 , (a=1,2,3) $ are the generators for the $SU(2)$ gauge group and $G^m, (m= 1, \dots , 8 )$ are the generators for the $SU(3)$ gauge group.  The two $\lambda_S$ singlet interacting terms will force VEVs for $\Phi_{2,3}$ and $\ov{\Phi}_{2,3}$. The first Yukawa interaction term is not relevant for the gauge symmetry breaking but  it will align the VEVs of $\Phi_{2,3} $ and $\ov{\Phi}_{2,3}$ in the  singlet component field direction, i.e.
\beq
\langle \Phi_3 \rangle\hspace{-3pt}  &=&\hspace{-3pt}  \left( \begin{array}{ccccc}
f_3\hspace{-3pt} &  0&0 &0 &0 \\
0 & f_3\hspace{-3pt} & 0&0 &0\\
0&0 & f_3\hspace{-3pt}&0 &0
\end{array} \right),\,
\langle \overline \Phi_3
\rangle^T\hspace{-2pt} = \hspace{-3pt} \left( \begin{array}{ccccc}
{\overline f_3}\hspace{-3pt} &  0&0 &0 &0 \\
0 &{\overline f_3}\hspace{-3pt} & 0&0 &0\\
0&0 &{\overline f_3}\hspace{-3pt} &0 &0
\end{array} \right),
\nonumber \\
\langle \Phi_2 \rangle\hspace{-3pt}  &=&\hspace{-3pt}  \left( \begin{array}{ccccc}
0 &0 &0 &f_2 \hspace{-3pt}&0 \\
0 &0 &0 &0 &f_2\hspace{-3pt}
\end{array} \right),
\, ~ \langle{\overline  \Phi_2}
\rangle^T\hspace{-2pt} =\hspace{-3pt}  \left( \begin{array}{ccccc}
0 &0 &0 &{\overline f_2}\hspace{-3pt} &0 \\
0 &0 &0 &0 &{\overline f_2}\hspace{-3pt}
\end{array} \right)\hspace{-1pt} .
\label{PHIbar}
\eeq
We can check from Table~ \ref{tab:content} that these singlets' VEVs do not violate the  $H+ V + a T_{24}$ charge.
The VEVs break the large gauge group down into  the MSSM gauge group $SU(3)_C\times SU(2)_L \times U(1)_Y$ and their gauge couplings  are given by:
\beq \frac{1}{{g_{2,3}^2 }} = \frac{1}{{\hat g_{2,3}^2 }} +
\frac{1}{{\hat g_5^2 }}, \quad\quad  \frac{1}{{g_{Y}^2 }} =
\frac{1}{{\hat g_{1H}^2 }} + \frac{1}{{\hat g_{1V}^2 }} + \frac{1}{15{\hat g_5^2 }},
\label{coupling} \eeq
where $\hat g_i$ and $\hat g_5$ are the gauge couplings of the original
$SU(3)$, $SU(2)$, $U(1)_H$, $U(1)_V$ and $SU(5)$ gauge groups
respectively. For simplicity, we further assume $f_2 = \ov{f}_2$ and $f_3 = \ov{f}_3$, therefore this is a D-term flat minimum and it will not induce mass terms for the Higgs fields. The singlet field $S_{1,2}$ and adjoint fields $A_{1,2}$ will not gain VEVs in this scenario.  These terms give the following scalar potential
for the four link fields $\Phi_{2,3}$ and $\ov{\Phi}_{2,3}$:
\beq
V_{\Phi} &=& y_1^2 \left |\phi_3 \ov{\phi}_2 \right |^2  + \lambda_S ^2 \left| {\mathop{\rm Tr}\nolimits} ~ \phi _2 \ov{\phi }_2 -w_2^2 \right|^2  + \lambda_S ^2 \left| {\mathop{\rm Tr}\nolimits} ~ \phi _3 \ov{\phi}_3 - w_3^2 \right|^2  \nonumber \\ &+&   \mu_2^2   \left( {\mathop{\rm Tr}\nolimits} ~ \phi _2^{\dagger}  \phi _2+  {\mathop{\rm Tr}\nolimits} ~ \ov{\phi} _2 \ov{\phi}^{\dagger} _2 \right)  + \mu_3^2  \left( {\mathop{\rm Tr}\nolimits} ~ \phi _3^{\dagger}  \phi _3+  {\mathop{\rm Tr}\nolimits} ~ \ov{\phi} _3 \ov{\phi}^{\dagger} _3 \right) \nonumber \\ & + & \lambda _A^2 {\left| { {\mathop{\rm Tr}\nolimits}~ { \ov{\phi} _2} ~ ( \sigma^a /2 ) ~ {\phi }_2} \right|^2 + \lambda_A^2 \left |{ {\mathop{\rm Tr}\nolimits} ~ {\ov{\phi} _3} {\rm G}^m {\phi }_3 }\right|^2}  \label{potential}
\eeq
The minimum of this simple potential determines the VEVs,
\beq
f_2^2 = \bar f_2^2 = \frac{ \lambda_s^2 w_2^2  - \mu_2^2 }{2 \lambda_s^2 }  ,  \label{VEV2} \\
f_3^2 = \bar f_3^2 = \frac{ \lambda_s^2 w_3^2  - \mu_3^2 }{3 \lambda _s^2 } .
\label{VEV3} \eeq
Substituting Eq.~[\ref{VEV2}] and Eq.~[\ref{VEV3}] back into the
scalar potential Eq.~[\ref{potential}], we can see that
Supersymmetry is spontaneously broken in this setup  via  the
ORaifeartaigh  mechanism with the simultaneous presence of  the
supersymmetric $\mu$ terms and the $\lambda_S$ interaction terms .
An easy way to verify this statement is that,  only if
$\mu_2=\mu_3=0$,  the  superpotential could have zero vacuum
energy when link fields develop nonzero VEVs.  Desired values for
the two VEVs $f_2 $ and $f_3$ can be achieved by tuning the three
free parameters: $\mu_{2,3}$, $ w_{2,3} $ and $\lambda_S$ .

As we expect no light modes  after the gauge symmetry breaking,
we first examine the mass spectrum in the link fields after  they
gain VEVs.   Due to the traceless properties of the $SU(2)$ and
$SU(3)$ gauge generators, the two $\lambda_A $ terms will not
change the VEVs, but they will give masses to the two linear
copies of real triplet fields  i.e.  ${\psi _{2T, 2}} =
\frac{1}{{\sqrt 2 }}{\mathop{\rm Re}\nolimits} \left( {{\phi
_{2T}} + \bar \phi _{2T}^\dag } \right)$ and  ${\psi _{2T, 3}} =
\frac{1}{{\sqrt 2 }}{\mathop{\rm Im}\nolimits} \left( {{\phi
_{2T}} - \bar \phi _{2T}^\dag } \right)$, as well as the two
linear copies of real octet fields i.e. ${\psi _{3O,2}} =
\frac{1}{{\sqrt 2 }}{\mathop{\rm Re}\nolimits} \left( {{\phi
_{3O}} + \bar \phi _{3O}^\dag } \right)$ and  ${\psi _{3O,3}} =
\frac{1}{{\sqrt 2 }}{\mathop{\rm Im}\nolimits} \left( {{\phi
_{3O}} - \bar \phi _{3O}^\dag } \right)$, leaving other states
untouched. The two $\lambda_S$ quartic terms and the two $
\mu_{2,3}^2$  mass terms can give masses to  two specific linear
copies of real triplet fields i.e.  ${\psi _{2T, 1}} =
\frac{1}{{\sqrt 2 }}{\mathop{\rm Re}\nolimits} \left( {{\phi
_{2T}} - \bar \phi _{2T}^\dag } \right)$ and  ${\psi _{2T, 3}} =
\frac{1}{{\sqrt 2 }}{\mathop{\rm Im}\nolimits} \left( {{\phi
_{2T}} - \bar \phi _{2T}^\dag } \right)$, plus  two specific
linear copies of  real octet i.e. ${\psi _{3O,1}} =
\frac{1}{{\sqrt 2 }}{\mathop{\rm Re}\nolimits} \left( {{\phi
_{3O}} - \bar \phi _{3O}^\dag } \right)$ and  ${\psi _{3O,3}} =
\frac{1}{{\sqrt 2 }}{\mathop{\rm Im}\nolimits} \left( {{\phi
_{3O}} - \bar \phi _{3O}^\dag } \right)$. They also give masses
for six real singlet states: ${\psi _{2S,1}} = \frac{1}{{\sqrt 2
}}{\mathop{\rm Re}\nolimits} \left( {{\phi _{2S}} - \bar \phi
_{2S}^\dag } \right)$, $ {\psi _{2S,2}} = \frac{1}{{\sqrt 2
}}{\mathop{\rm Re}\nolimits} \left( {{\phi _{2S}} + \bar \phi
_{2S}^\dag } \right)$,  ${\psi _{2S,3}} = \frac{1}{{\sqrt 2
}}{\mathop{\rm Im}\nolimits} \left( {{\phi _{2S}} - \bar \phi
_{2S}^\dag } \right) $; ${\psi _{3S,1}} = \frac{1}{{\sqrt 2
}}{\mathop{\rm Re}\nolimits} \left( {{\phi _{3S}} - \bar \phi
_{3S}^\dag } \right)$, ${\psi _{3S,2}} = \frac{1}{{\sqrt 2
}}{\mathop{\rm Re}\nolimits} \left( {{\phi _{3S}} + \bar \phi
_{3S}^\dag } \right)$, $ {\psi _{3S,3}} = \frac{1}{{\sqrt 2
}}{\mathop{\rm Im}\nolimits} \left( {{\phi _{3S}} - \bar \phi
_{3S}^\dag } \right)$. Ignoring some singlets mixing,  we list
the mass spectrum for all the singlets,  triplets and octets  in
Table~\ref{scalarmass}.  As shown in that table, we have one copy
of real triplet field , one copy of real octet field and two
copies of real singlet fields left massless,  which are the
Goldstone bosons eaten by the heavy $W '$, $G ' $ gauge bosons and
two heavy  $U(1) $ gauge bosons  $B'$ and $B''$ respectively.
\begin{table}
\begin{center}
\begin{tabular}{ccccc} \hline \\
${\frac{1}{{\sqrt 2 }} {\mathop{\rm Re}}\left( {{\phi _{2T}} - \bar \phi _{2T}^\dag } \right)} $ :  & $ { 2\mu _2^2 + 2 ( \hat g_2^2+\hat g_5^2) f_2^2}$  & $ \qquad  \qquad $ & $ {\frac{1}{{\sqrt 2 }} {\mathop{\rm Re}}\left( {{\phi _{2T}} + \bar \phi _{2T}^\dag } \right)} $ :   &   $\lambda_A^2 f_2^2$
\\ \\
${\frac{1}{{\sqrt 2 }} {\mathop{\rm Im}} \left( {{\phi _{2T}} - \bar \phi _{2T}^\dag } \right)} $ :  & $ { \left( 2 \mu _2^2 + \lambda_A^2 f_2^2 \right) }$  & $ \qquad  \qquad $ & $ {\frac{1}{{\sqrt 2 }} {\mathop{\rm Im}} \left( {{\phi _{2T}} + \bar \phi _{2T}^\dag } \right)} $ :   & 0
\\ \\
${\frac{1}{{\sqrt 2 }}{\mathop{\rm Re}} \left( {{\phi _{2S}} - \bar \phi _{2S}^\dag } \right)}$ : &  $ {2 \mu _2^2 } $
& $\qquad \qquad $ & $ {\frac{1}{{\sqrt 2 }}{\mathop{\rm Re}} \left( {{\phi _{2S}} + \bar \phi _{2S}^\dag } \right)} $ : & ${\left(2 \lambda _S^2 w_2^2 - 2 \mu_2^2 \right)} $ \\ \\
 ${\frac{1}{{\sqrt 2 }}{\mathop{\rm Im }} \left( {{\phi _{2S}} - \bar \phi _{2S}^\dag } \right)} $ : & $ { 2 \lambda _S^2 w_2^2} $
& $\qquad \qquad$  & ${\frac{1}{{\sqrt 2 }}{\mathop{\rm Im}} \left( {{\phi _{2S}} + \bar \phi _{2S}^\dag} \right)} $ : & $0 $
\\ \\
$ {\frac{1}{{\sqrt 2 }} {\mathop{\rm Re}}\left( {{\phi _{3O}} - \bar \phi _{3O}^\dag } \right)}$ : &  ${ 2 \mu _3^2 +2 (\hat g_3^2+ \hat g_5^2) f_3^2} $ & $ \qquad \qquad $ & $ {\frac{1}{{\sqrt 2 }} {\mathop{\rm Re}}\left( {{\phi _{3O}} + \bar \phi _{3O}^\dag} \right)} $ &    $ \lambda_A^2 f_3^2$
\\ \\
$ {\frac{1}{{\sqrt 2 }} {\mathop{\rm Im}}\left( {{\phi _{3O}} - \bar \phi _{3O}^\dag } \right)}$: &  ${ \left( 2 \mu _3^2 +  \lambda_A^2 f_3^2 \right) } $ & $ \qquad \qquad $ & $ {\frac{1}{{\sqrt 2 }} {\mathop{\rm Im}}\left( {{\phi _{3O}} + \bar \phi _{3O}^\dag } \right)} $ & $ 0 $
\\ \\
${\frac{1}{{\sqrt 2 }}{\mathop{\rm Re}} \left( {{\phi _{3S}} - \bar \phi _{3S}^\dag } \right)} $ & $ { 2 \mu _3^2}  $ & $\qquad \qquad$ &  ${\frac{1}{{\sqrt 2 }}{\mathop{\rm Re}} \left( {{\phi _{3S}} + \bar \phi _{3S}^\dag } \right)}$ & $ {\left(2 \lambda _S^2 w_3^2 - 2 \mu_3^2 \right)}$  \\ \\
${\frac{1}{{\sqrt 2 }}{\mathop{\rm Im}} \left( {{\phi _{3S}} - \bar \phi _{3S}^\dag } \right)} $ & $ {2 \lambda _S^2w_3^2} $ & $\qquad \qquad $ & $ {\frac{1}{{\sqrt 2 }}{\mathop{\rm Im}} \left( {{\phi _{3S}} +\bar \phi _{3S}^\dag } \right)} $ & $0$  \\ \\ \hline
\end{tabular}  \nonumber
\end{center}
\label{scalarmass}
\caption{ mass spectrum for singlets ( ignore their mixing ), triplets and octets in the  link fields $\phi_{2,3}$ and $\ov{\phi}_{2,3}$.}
\end{table}

The two real singlets fields $\psi_{2S,1} = \frac{1}{\sqrt 2} \mathop{\rm Re} \left( {{\phi _{2S}} - \bar \phi _{2S}^\dag } \right)$  and $ \psi_{3S, 1} = \frac{1}{\sqrt 2} \mathop{\rm Re} \left( {{\phi _{3S}} - \bar \phi _{3S}^\dag } \right)$  mix in a more complicated way due to $U(1)$ D-terms proportional to $\hat g_5^2 $, $\hat g_{1H}^2 $ and $\hat g_{1V}^2$. Their mass eigenstates are determined by diagonalizing the full mass terms for $\psi_{2S,1}$ and $\psi_{3S,1}$ as described in Eq.~[\ref{singlets}]:
\beq
\frac{{2 \hat g_5^2}}{{15}} {\left( {3  {f_2}{\psi _{2S,1}} - \sqrt 6 {f_3}{\psi _{3S,1}}} \right)^2} + \frac{{\hat g_{1H}^2}}{3} f_3^2\psi _{3S,1}^2 + \frac{{\hat g_{1V}^2}}{{50}} {\left( { 2 {f_2}{\psi _{2S,1}} + \sqrt 6 {f_3}{\psi _{3S,1}}} \right)^2} + 2\mu _2^2{\psi _{2S,1}^2} + 2 \mu _3^2{\psi _{3S,1}^2}\label{singlets}
\eeq

The component fields in the off-diagonal direction also need to be
dealt with in a straightforward way. For convenience, we conduct
an eigenstate basis rotation and  redefine those stop-like states
in the four link fields as follows:
\beq
{\psi _{2t, 1}} = \frac{1}{{\sqrt 2 }}{\mathop{\rm Re}} ({\phi _{2t}} - \bar \phi _{2t}^\dag ) , ~
{\psi _{2t, 2}} = \frac{1}{{\sqrt 2 }}{\mathop{\rm Re}} ({\phi _{2t}} + \bar \phi _{2t}^\dag ) , ~
{\psi _{2t, 3}} = \frac{1}{{\sqrt 2 }}{\mathop{\rm Im}} ({\phi _{2t}} - \bar \phi _{2t}^\dag ) , ~
{\psi _{2t, 4}} = \frac{1}{{\sqrt 2 }}{\mathop{\rm Im}} ({\phi _{2t}} + \bar \phi _{2t}^\dag ) ,
\\
{\psi _{3t,1}} = \frac{1}{{\sqrt 2 }} {\mathop{\rm Re}} ({\phi _{3t}} - \bar \phi _{3t}^\dag ) , ~
{\psi _{3t,2}} = \frac{1}{{\sqrt 2 }} {\mathop{\rm Re}} ({\phi _{3t}} + \bar \phi _{3t}^\dag ) , ~
{\psi _{3t,3}} = \frac{1}{{\sqrt 2 }} {\mathop{\rm Im}} ({\phi _{3t}} - \bar \phi _{3t}^\dag ) , ~
{\psi _{3t,4}} = \frac{1}{{\sqrt 2 }} {\mathop{\rm Im}} ({\phi _{3t}} + \bar \phi _{3t}^\dag ) .
\eeq
In terms of the new eigenstate basis of $\psi_{2t, i}, i =1,2,3, 4 $,  and $ \psi_{3t, i }, i = 1,2 ,3, 4 $, the mass terms from both the gauge interactions and superpotentials are:
\beq
&& \hat g_5^2{({f_2}{\psi _{2t,1}} + {f_3}{\psi ^{T}_{3t,1}})^2} + 2\mu _2^3\psi _{2t,1}^2 + 2\mu _3^3(\psi^T _{3t,1})^2 + \frac{{y_1^2}}{2}{({f_2}({\psi ^T_{3t,1}} + {\psi^T _{3t,2}}) - {f_3}({\psi _{2t,1}} - {\psi _{2t,2}}))^2} ~ ,  \label{stop1} \\
&& \hat g_5^2{({f_2}{\psi _{2t,3}} - {f_3}{\psi ^T_{3t,3}})^2} + 2\mu _2^3\psi _{2t,3}^2 + 2\mu _3^3(\psi^T _{3t,3})^2 + \frac{{y_1^2}}{2}{({f_2}({\psi ^T_{3t,3}} + {\psi^T _{3t,4}}) + {f_3}({\psi _{2t,3}} - {\psi _{2t,4}}))^2} ~ .\label{stop2}
\eeq
The origins of each term in the above two equations are:  the
first one is  from the Super-Higgs mechanism,  the second  one and
third one are from  the $\mu_{2,3}$ mass terms as well as the
$\lambda_S$ interaction terms, while the last one comes from the
$y_1 Q_3 \Phi_2 \ov{\Phi}_3$ Yukawa interaction. Similar to the
singlets case, the exact mass eigenstates are determined by
diagonalizing these two equations.  Examining Eq.~[\ref{stop1}] and
Eq.~[\ref{stop2}], it is easy to find out that only two linear
combinations of stop-like  states are  still massless and they
should be identified as the Goldstone bosons for the heavy $X$, $Y$
gauge bosons.
\beq
\pi_t &=& \frac{1}{{\sqrt {f_2^2 + f_3^2} }}({f_2}{\psi _{2t,2}} - {f_3}{\psi^T _{3t,2}})\\
\eta_t  &=& \frac{1}{{\sqrt {f_2^2 + f_3^2} }}({f_2}{\psi _{2t,4}} + {f_3}{\psi^T _{3t,4}})
\eeq

With the mass spectrum for all the scalars fields in  $\phi_{2,3}
$ and $\ov{\phi}_{2,3}$,  we proceed to discuss the effective
D-term in this model. In a supersymmetric gauge theory, when a
large gauge group breaks down into the MSSM gauge group, the
vector supermultiplet corresponding to the unbroken generators
will inherit the MSSM gauge interactions and remain massless
after the gauge symmetry breaking.  For the vector supermultiplet
corresponding to the broken generators, they could achieve masses
after eating a copy of chiral supermultiplet through the
Super-Higgs mechanism. In the supersymmetric limit, all component
fields $( A_\mu, \lambda_1, \lambda_2, \Sigma ) $ in a heavy
vector supermultiplet have degenerate masses. And after
integrating them out , the D-term effects from those heavy states
will decouple . In order to retain the D-term effects from those
heavy states till the low energy scale, a SUSY breaking mass term
need to be added to the real scalar component field $\Sigma$,
i.e. the lowest component field in the heavy vector
supermultiplet,  which will recouple the D-term effects from the
broken gauge generators back into the effective Lagrangian.  In
the low energy scale, since the Higgs bosons  are charged under
the diagonal $SU(2)_W $ and $U(1)_Y$ gauge group, there should be
two sources of D-term enhancement in this model: One is from an
extra $SU(2)$ embedded in the $SU(5)$ gauge group and the other is
from two extra $ U(1)$s.  For the heavy $SU(2)$ vector
supermultiplet,  its corresponding scalar component field is  a
triplet state : $\psi_{2T, 1} = \frac{1}{\sqrt 2} \mathop{\rm Re}
\left( {{\phi _{2T}} - \bar \phi _{2T}^\dag } \right)$. While for
two extra $U(1)$s ,  their scalar component fields  are  two heavy
singlets: $\psi_{2S, 1} =\frac{1}{\sqrt 2} \mathop{\rm Re} \left(
{{\phi _{2S}} - \bar \phi _{2S}^\dag } \right)$ and $ \psi_{3S, 1}
= \frac{1}{\sqrt 2} \mathop{\rm Re} \left( {{\phi _{3S}} - \bar
\phi _{3S}^\dag } \right)$.  After integrating out those heavy
fields, an effective D-term  is obtained for the Higgs bosons :
\beq
\frac{g_2^2}{2} \Delta_2 \left( H_2^{\dagger
   }\frac{\sigma ^a}{2} H_2 - H_1^{\dagger } \frac{\sigma ^a }{2} H_1\right)^2+\frac{g_{Y}^2}{2} \Delta_Y
    \left(\frac{1}{2} H_2^{\dagger } H_2-\frac{1}{2}
   H_1^{\dagger } H_1\right)^2 ~,
\eeq
where $g_2 $ is the gauge coupling for the SM gauge group
$SU(2)_W$ and $g_Y $ is the gauge coupling for  the SM hypercharge
gauge group $U(1)_Y$, whose values are determined by
Eq.~[\ref{coupling}] . The D-term effects of these heavy scalar
fields are nondecoupling due to spontaneous SUSY breaking effects
in our scenario, and their  effects can be summarized  into two
parameters $ \Delta_2$ and $\Delta_Y$:
\beq
\Delta_2 &=& \left({1+\frac{  m_{2T}^2 }{f_2^2 }\frac{1}{2 \hat{g}_2^2}}\right) / \left( {1+\frac{ m_{2T}^2 }{f_2^2 }\frac{1}{2 \left(\hat{g}_5^2+\hat{g}_2^2\right)}} \right)  ~,
\eeq
\beq
\Delta_Y &=&  \left( 1 + \left( {\frac{8}{15\hat g_5^2} + \frac{1}{2\hat g_{1H}^2} + \frac{1}{2\hat g_{1V}^2}} \right)\frac{ m_{2S}^2}{f_2^2} + \left( {\frac{1}{5\hat g_5^2} + \frac{3}{ \hat g_{1V}^2}} \right)\frac{ m_{3S}^2}{f_3^2}  \right.  \nonumber \\  &+& \left.  \frac{15\hat g_5^2 + 25\hat g_{1H}^2 + \hat g_{1V}^2}{10\hat g_5^2\hat g_{1H}^2\hat g_{1V}^2}\frac{ m_{2S}^2m_{3S}^2}{f_2^2f_3^2} \right) / \left( 1 + \frac{60\hat g_5^2 + 25\hat g_{1H}^2 + 9\hat g_{1V}^2}{2(\hat g_{1H}^2\hat g_{1V}^2 + 15\hat g_5^2(\hat g_{1H}^2 + \hat g_{1V}^2)) }\frac{ m_{2S}^2}{f_2^2}  \nonumber \right. \\  &+&  \left. \frac{3(15\hat g_5^2 + \hat g_{1V}^2)}{ ( \hat g_{1H}^2\hat g_{1V}^2 + 15\hat g_5^2(\hat g_{1H}^2 + \hat g_{1V}^2))}\frac{ m_{3S}^2}{f_3^2} + \frac{75}{2(\hat g_{1H}^2\hat g_{1V}^2 + 15\hat g_5^2(\hat g_{1H}^2 + \hat g_{1V}^2))}\frac{ m_{2S}^2m_{3S}^2}{f_2^2f_3^2} \right) ~ . \eeq
$m_{2T}$, $m_{2S} $ and $m_{3S}$ are the respective F-term masses
for  the heavy scalar fields $\psi_{2T,1} $, $\psi_{2S,1} $ and
$\psi_{3S,1}$  induced by spontaneous SUSY breaking , whose
values can be read from the first term in each column of Table~
\ref{scalarmass}:
\beq  m_{2T}^2 = {2 \mu _2^2 } ~,  \qquad m_{2S}^2 =  {2 \mu _2^2 } ~, \qquad  m_{3S}^2 = {  2 \mu _3^2 }  \eeq
In the supersymmetric limit  i.e.  $\mu_ 2 =  \mu_3 =0$,   we can find that $\Delta_2 =1$ and $\Delta_Y=1$,  that is the D-term effects from those heavy fields  are decoupled . However in this model, since we prefer to stay in the region of $ \hat g_5 f_3 \ll  \mu_3 $ and  $ \hat g_5 f_2 \ll  \mu_2 $,  we expect that there are notable enhancements for both of the $SU(2)_W$ and $U(1)_Y$  D-terms.  Under the limit of $ m_{2T} \gg f_2 $,  $ m_{2S} \gg  f_2$, and $ m_{3S}  \gg  f_3$,  $\Delta_2 $ and $\Delta_Y $ are simply determined by those gauge coupling constants:
\beq   {\Delta _2} \approx \frac{\hat g_2^2 + \hat g_5^2}{\hat g_2^2 ~ \hat g_5^2}  \hat g_5^2 = \frac{ \hat g_5^2 }{g_2^2} ~,  \qquad \qquad   {\Delta _Y} \approx  \frac{15\hat g_5^2 + 25\hat g_{1H}^2 + \hat g_{1V}^2 }{25g_Y^2}  \eeq
We can see that in the large SUSY breaking limit, the effective D-term for the Higgs bosons is only proportional to three gauge coupling constants $\hat g_5^2$, $\hat g_{1H}^2$ and $\hat g_{1V}^2 $, which  is exactly the same as in the original unbroken gauge theory. The Higgs bosons can gain notable mass through the D-term enhancement effects as long as we choose the gauge couplings under which our Higgs bosons are charged to be larger than the gauge couplings in the other moose site.  An simple example is  choosing $\hat g_2 = 0.78$,  $ \hat g_5 = 1.2$, $\hat g_{1H} = 0.378 $ and $ \hat g_{1V}$ =1.5, we will obtain  $ \Delta_2 \sim 3.36 $ and $ \Delta_Y \sim 8.21 $. With an $O(1) $   $\tan(\beta) \sim 2.0$, at the tree level the Higgs mass squared $ M_h^2 \le \frac{1}{4} (g_2^2 \Delta_2 + g_Y^2 \Delta_Y ) v^2 \cos^2(2 \beta) $ can be naturally raised to be around $ (115 ~ \mbox{GeV})^2$. Since the radiation corrections from the top quark and its superpartners will contribute to the running of the Higgs quartic term:
\beq
\delta \lambda  = \frac{3\hat g_5^4}{8 {\pi ^2}}\ln \left( \frac{ m_{
\vec Q } m_{{\tilde t}_R} }{ m_t^2 } \right) ~,
\eeq
at the loop level the Higgs mass is further enhanced  such that
\beq
&& m_h^2 = \frac{1}{2}\left( {m_A^2 + M_G^2- \sqrt {\left( {m_A^2 + M_G^2}\right)^2 - 4 m_A^2 M_G^2 {{\cos }^2}2\beta }} \right) + 2 \delta \lambda v^2 \sin(\beta)^2 ~ ,  \label{Higgs}\\
&& \mbox{with}    \qquad \qquad  M_G^2 =   \frac{1}{4}\left( {g_2^2{\Delta _2} + g_Y^2{\Delta _Y}} \right){v^2} \nonumber  ~.
\eeq
The precision of Higgs mass prediction  depends on  the  mass of  vector top partner and  the mass of the right handed stop  used to calculate the radiation correction.  The vector top partner gains its mass through the link fields' VEVs, i.e. $m_{\vec Q} =  \hat g_5^2 ( f_2 ^2 + f_3^2 ) $. While the right handed stop acquires its mass through the higgs $\mu_H$ term $\mu_H H \bar H $ as well as from the soft SUSY breaking scalar mass term.  If  the vector top partner mass is $m_{ \vec Q }\simeq 2.8 ~\mbox{TeV} $  and  the mass of right handed stop is  $m_{{\tilde t}_R} \simeq 300  ~\mbox{GeV}$,  and set the mass parameter to be $M_A = 800~\mbox{GeV }$,  we can get a heavy Higgs bosons $m_h  \simeq 195 ~ \mbox{GeV} $.

\section{Electroweak Constraints: $S$, $T$, $U$ and $Z \to b \bar b$ }

Those gauge couplings should be chosen so that they could
reproduce the SM gauge couplings at the EW scale after running by
the renormalization group . In the following, we are going to take
some specific sets  of gauge couplings when we proceed with the
electroweak analysis, so that the electroweak measurements are
adopted to constrain  the link field VEVs, $f_2$ and $f_3$, whose
values in turn determine the masses of the vector top partner and
Higgs boson in this model.  The most stringent constraints come
from the mixing of $W$, $B$ and $W'$, $B'$, $B''$ due to the Higgs
VEVs because the two Higgs doublets and the light fermions
transform under different gauge groups. Oblique corrections to the
Standard Model are contained in the vacuum polarizations  of gauge
bosons, which are parameterized by $S$, $T$ and $U$. The vacuum
polarizations of a gauge boson can be expanded around the zero
momentum.
\beq \Pi _{a,a'} (p^2 ) = \Pi _{a,a'} (0) + p^2 \Pi '_{a,a'} (0) +
\cdots~, \eeq
and the $S$,  $T$ and $U $ parameters are defined in the following way:
\begin{eqnarray}
S = 16 \pi \cdot (\Pi '_{33} (0)-\Pi '_{3Q}), \qquad  T =
\frac{4 \pi}{s^2 c^2 M_Z^2} (\Pi _{11} (0) - \Pi _{33} (0)), \qquad  U =16 \pi \cdot (\Pi '_{11} (0)-\Pi '_{33}).
\label{STU}\end{eqnarray}
with $s= \sin(\theta )$ and $ c =\cos(\theta) $ and $\theta$ is
Weinberg angle. The definition of $S, T, U $ subtracts out the
predicted SM contribution with fixed top quark mass and Higgs
bosons mass so that they  encode only  new physics
contributions.The contribution to the $S$ and $U$ parameters from
the gauge bosons mixing is very less  , which are at the order of
$ v^4/f_3^4 $ or $ v^4/f_2^4$. The analytic expressions for $S$
and $U$ are simply given by:
\beq \Delta S &=& - \frac{\pi }{4} \left( \frac{{ \left( {25\hat g_{1H}^2 + 60\hat g_5^2 + 9\hat g_{1V}^2} \right)\left( {15\hat g_5^2\hat g_{1V}^2 + \hat g_{1H}^2\left( { - 15\hat g_5^2 + 4\hat g_{1V}^2} \right)} \right)^2}}{{{{\left( {15\hat g_5^2\hat g_{1V}^2 + \hat g_{1H}^2\left( {15\hat g_5^2 + \hat g_{1V}^2} \right)} \right)}^3}}}\frac{{{v^4}}}{{f_2^4}} + \frac{{\hat g_5^4}}{{{{\left( {\hat g_2^2 + \hat g_5^2} \right)}^3}}}\frac{{{v^4}}}{{f_2^4}} \right) \nonumber \\ & & -   \frac{{9\pi \hat g_{1H}^2\left( {150\hat g_5^4 - 5\hat g_5^2\hat g_{1V}^2 - \hat g_{1V}^4} \right)\left( { - 15\hat g_5^2\hat g_{1V}^2 + \hat g_{1H}^2\left( {15\hat g_5^2 - 4\hat g_{1V}^2} \right)} \right)}}{{\left( {15\hat g_5^2\hat g_{1V}^2 + \hat g_{1H}^2\left( {15\hat g_5^2 + \hat g_{1V}^2} \right)} \right)^3}}\frac{{{v^4}}}{{f_2^2f_3^2}}  \nonumber \\ & & -  \frac{{9\pi \hat g_{1H}^4{{\left( {15\hat g_5^2 + \hat g_{1V}^2} \right)}^3}}}{{{{\left( {15\hat g_5^2\hat g_{1V}^2 + \hat g_{1H}^2\left( {15\hat g_5^2 + \hat g_{1V}^2} \right)} \right)}^3}}}  \frac{v^4}{f_3^4}  ~, \\ \nonumber   \\ \nonumber  \\
\Delta U &=& \frac{\pi }{4} \frac{{ \left( {25\hat g_{1H}^2 + 60\hat g_5^2 + 9\hat g_{1V}^2} \right)\left( {15\hat g_5^2\hat g_{1V}^2 + \hat g_{1H}^2\left( { - 15\hat g_5^2 + 4\hat g_{1V}^2} \right)} \right)^2}}{{{{\left( {15\hat g_5^2\hat g_{1V}^2 + \hat g_{1H}^2\left( {15\hat g_5^2 + \hat g_{1V}^2} \right)} \right)}^3}}}\frac{{{v^4}}}{{f_2^4}}  \nonumber \\ & & +  \frac{{9\pi \hat g_{1H}^2\left( {150\hat g_5^4 - 5\hat g_5^2\hat g_{1V}^2 - \hat g_{1V}^4} \right)\left( { - 15\hat g_5^2\hat g_{1V}^2 + \hat g_{1H}^2\left( {15\hat g_5^2 - 4\hat g_{1V}^2} \right)} \right)}}{{\left( {15\hat g_5^2\hat g_{1V}^2 + \hat g_{1H}^2\left( {15\hat g_5^2 + \hat g_{1V}^2} \right)} \right)^3}}\frac{{{v^4}}}{{f_2^2f_3^2}}  \nonumber \\ & & +  \frac{{9\pi \hat g_{1H}^4{{\left( {15\hat g_5^2 + \hat g_{1V}^2} \right)}^3}}}{{{{ \left( {15\hat g_5^2\hat g_{1V}^2 + \hat g_{1H}^2\left( {15\hat g_5^2 + \hat g_{1V}^2} \right)} \right)}^3}}}  \frac{v^4}{f_3^4}  ~.
\eeq
where  it is easy to  verify  that $\Delta S$ and $\Delta U$ are related by the following  identify:
\beq
\Delta S &=& - \Delta U - \frac{\pi}{4}  \frac{{\hat g_5^4}}{{{{\left( {\hat g_2^2 + \hat g_5^2} \right)}^3}}}\frac{{{v^4}}}{{f_2^4}} ~.
\eeq
The experimental constraints for $S$, $T$ and $U$ are given by~\cite{EW}:
\beq
S= 0.04 \pm 0.10  ~,  \qquad  T = 0.05 \pm 0.12 ~, \qquad   U = 0.08 \pm 0.11 ~.
\eeq
If we assume $0.4~\mbox{TeV} <  f_3 \ll f_2  $, and  with a small $\hat g_{1H} $ but a large $\hat g_{1V } $, the $S$ and $U$ parameters do not put any constraint to our parameter space. The situation is different for  the other oblique parameter. The gauge bosons mixing can give sizable contribution to the  $T$ parameter:
\beq
\Delta T = \frac{1}{\alpha }\left(\frac{\left( 15 \hat g_5^2 \hat g_{1V}^2 + \hat g_{1H}^2\left( - 15 \hat g_5^2 + 4 \hat g_{1V}^2\right) \right)^2}{\left( 15 \hat g_5^2 \hat g_{1V}^2 + \hat g_{1H}^2 \left( 15 \hat g_5^2 + \hat g_{1V}^2 \right) \right)^2}\frac{v^2}{8 f_2^2 } + \frac{3 \hat g_{1H}^4 \left(15 \hat g_5^2 + \hat g_{1V}^2 \right)^2}{\left(15 \hat g_5^2 \hat g_{1V}^2 + \hat g_{1H}^2 \left(15 \hat g_5^2 + \hat g_{1V}^2 \right) \right)^2}\frac{v^2}{4 f_3^2 } \right) ~.\label{TG}
\eeq
There is another big source of  $T$  parameter contribution  from
the heavy Higgs~\cite{peskin},  with the reference higgs mass of 
$m_{href} = 120~\mbox{GeV}$:
\beq
\Delta {T_h} =  - \frac{3}{{16 \pi c^2}}\log \left( \frac{m_h^2}{m_{href}^2} \right) ~. \label{TH}
\eeq
Since in the interested region of parameter space this model gives
a negligible contribution to the $U$ parameter,  we  can fix $U=0$,
therefore the experimental constraints for the $T$ parameter is
\cite{EW}:
\beq
T|_{U=0}=0.10 \pm 0.08 ~.
\eeq
\begin{figure}[htb]
\begin{center}
\includegraphics[width=0.48\hsize]{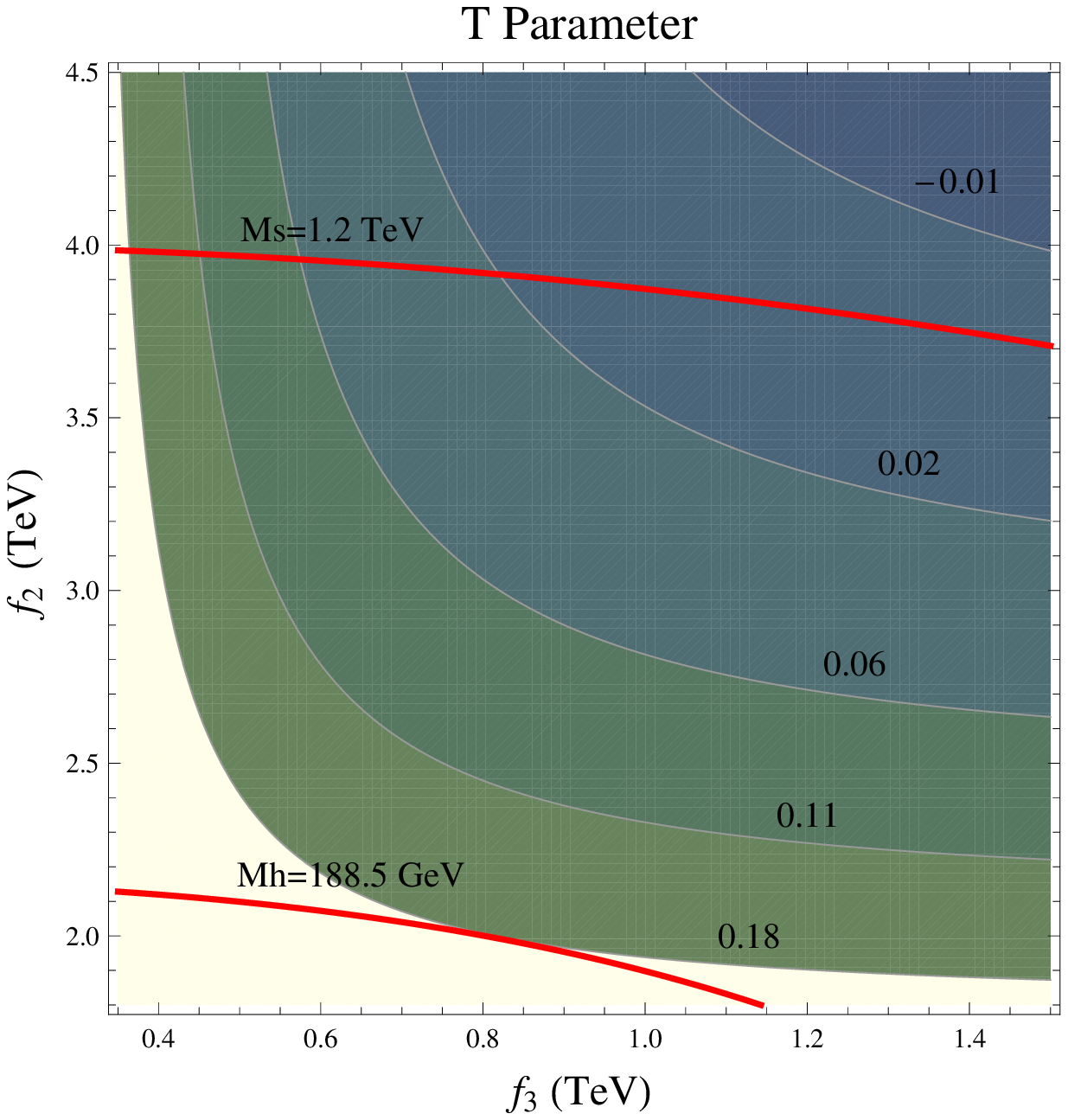}
\includegraphics[width=0.48\hsize]{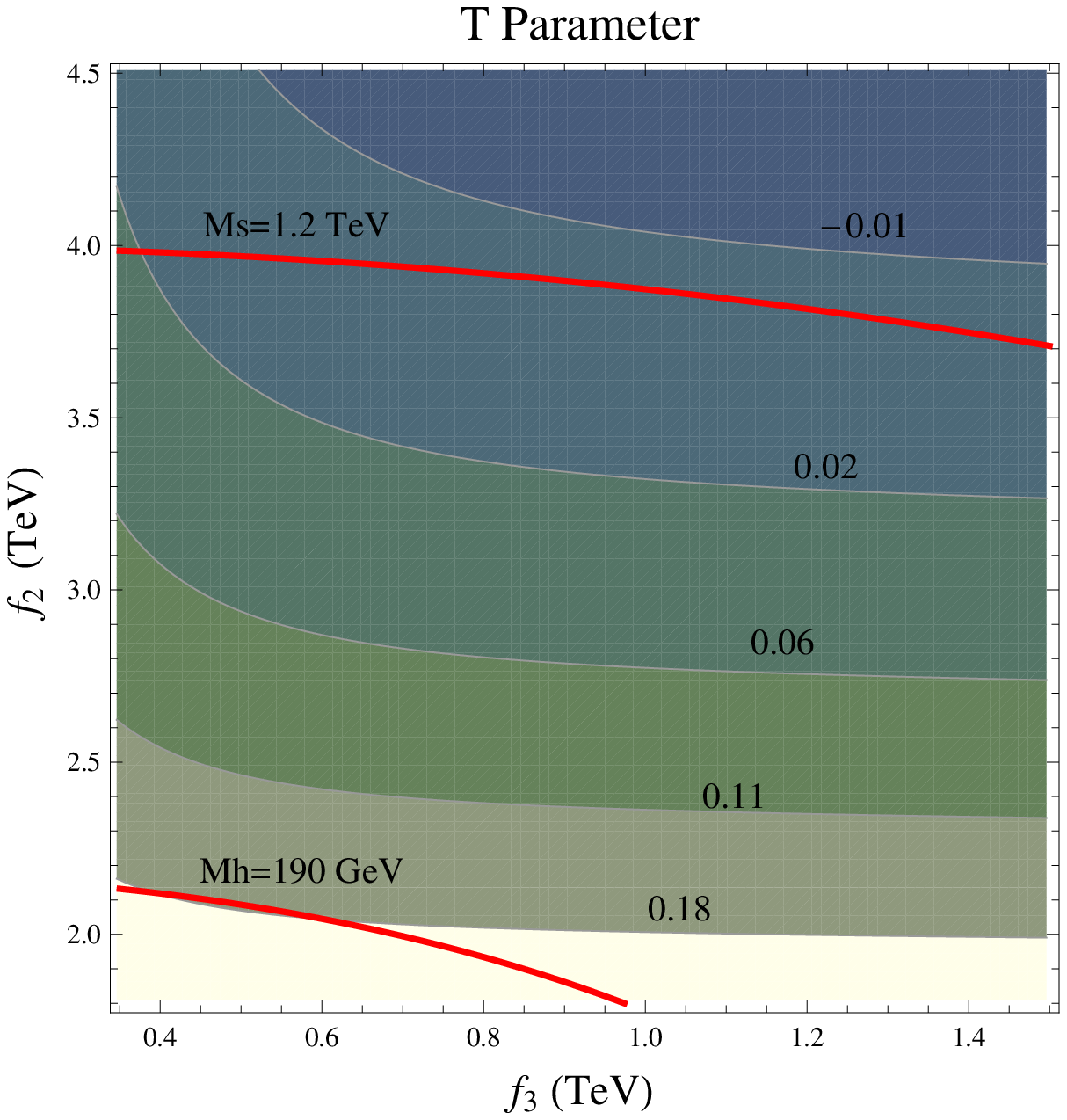}
\end{center}
\caption{ Contours of $T$ parameter as functions of two VEVs $f_2$ and $f_3$. Input parameters are: $ m_{\tilde t_R} = 300~\mbox{GeV}$,   $\tan \beta = 2.0 $, $\mu_2 = 5 ~\mbox{TeV}$, $\mu_3 = 2 ~\mbox{TeV} $, and  $ M_A = 800~\mbox{GeV} $.  In the left panel,  the gauge couplings are $\hat g_5 =1.2$, $\hat  g_{1H} =0.378$, $\hat g_{1V} =1.5$, and $\hat g_2=0.78$.  In the right panel, the gauge couplings are $ \hat g_{5} =1.2$, $\hat g_{1H}=0.37$, $ \hat g_{1V} =2.5$, and $\hat g_2=0.78$. The bottom red lines in both plots put a lower bound for the Higgs bosons mass under the requirement of $T < 0.18$.  The SUSY breaking scale $M_S = \sqrt {m_{\vec Q} m_{\tilde t_R} }$ is set to be $1.2$ TeV, corresponding to the top red lines in both plots,  which puts an upper bound for the parameter space. \label{HTPara}}
\end{figure}
\begin{table}
\begin{center}
\begin{tabular}{cccccccccccc} \hline  \\
${ m_{{\tilde t}_R} (\mbox{GeV}) } $ &  $ { 300  } $ &   & $ {350 } $&  & $ {400  } $ &  & $ {450 }$ & & $ {500 }$ &  & $ {550  } $
 \\ \\  \hline \\
$ {{m_h} (\mbox{GeV})} $ & $ 188.5  $&  & $ 191.5 $ & & $194.0  $ & & $ 196.5 $& & $ 198.5  $ & & $200 $
\\  \\ \hline \end{tabular}
\end{center}
\label{higgs}
\caption{ Input parameters are: $\mu_2 = 5 ~\mbox{TeV}$, $\mu_3 = 2 ~\mbox{TeV} $,  $M_A = 800 ~\mbox{GeV} $, and $\tan \beta = 2.0 $. Gauge couplings are taken to be $\hat g_5 =1.2$, $ \hat g_{1H} =0.378$, $\hat g_{1V} =1.5$ and $\hat g_2=0.78$.  Increasing the right handed stop mass from $ 300 ~\mbox{GeV} $ to $550 ~\mbox{GeV}$ in a modest way without inducing large fine tuning,   the respective lower bound for  the Higgs bosons mass can be calculated by requiring $T < 0. 18$. As we can see, $m_h$  varies from $188.5~\mbox{GeV}$ to $200~\mbox{GeV}$. }
\end{table}
\begin{figure}[htb]
\begin{center}
\includegraphics[width=0.48\hsize]{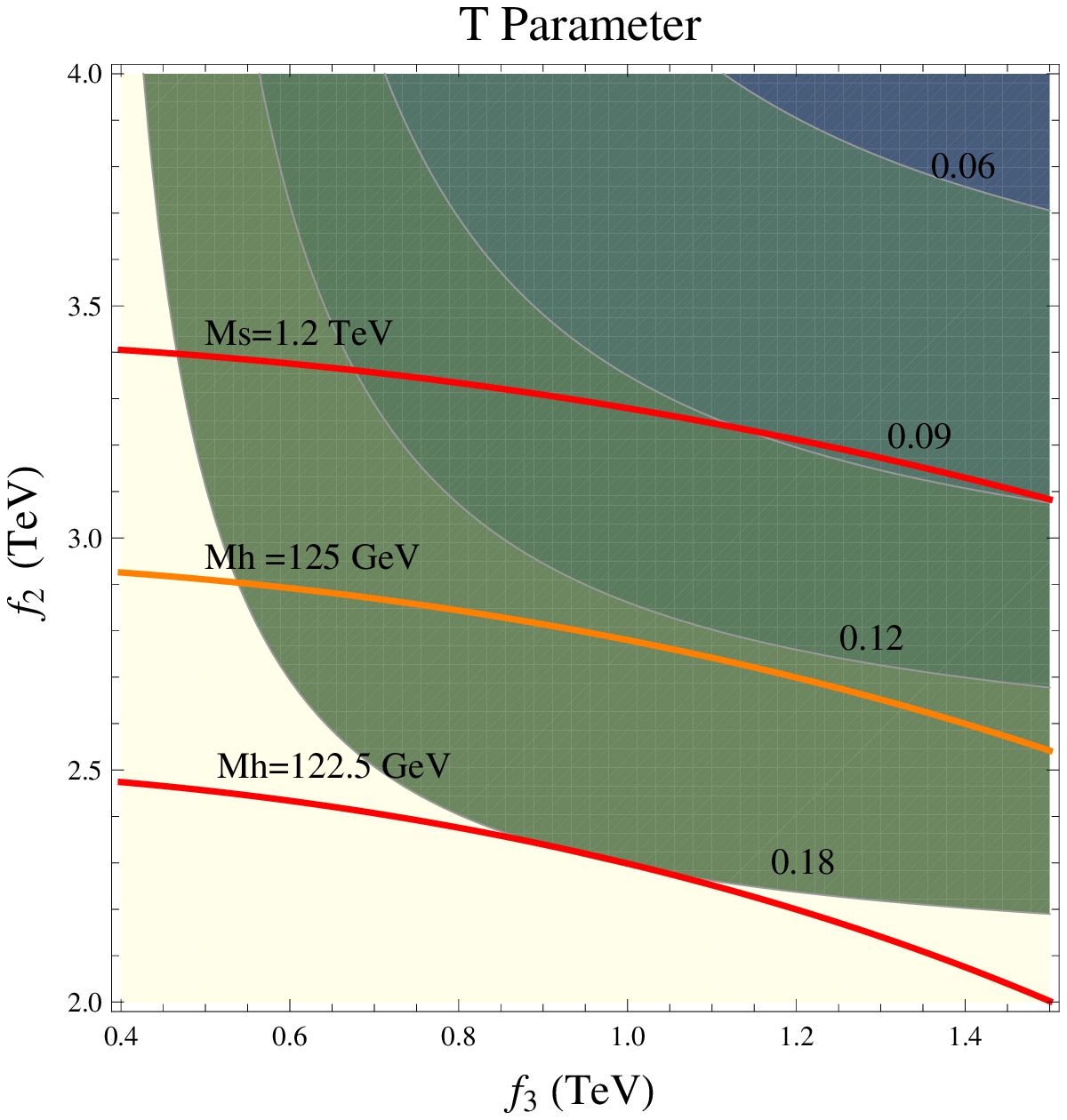}
\includegraphics[width=0.48\hsize]{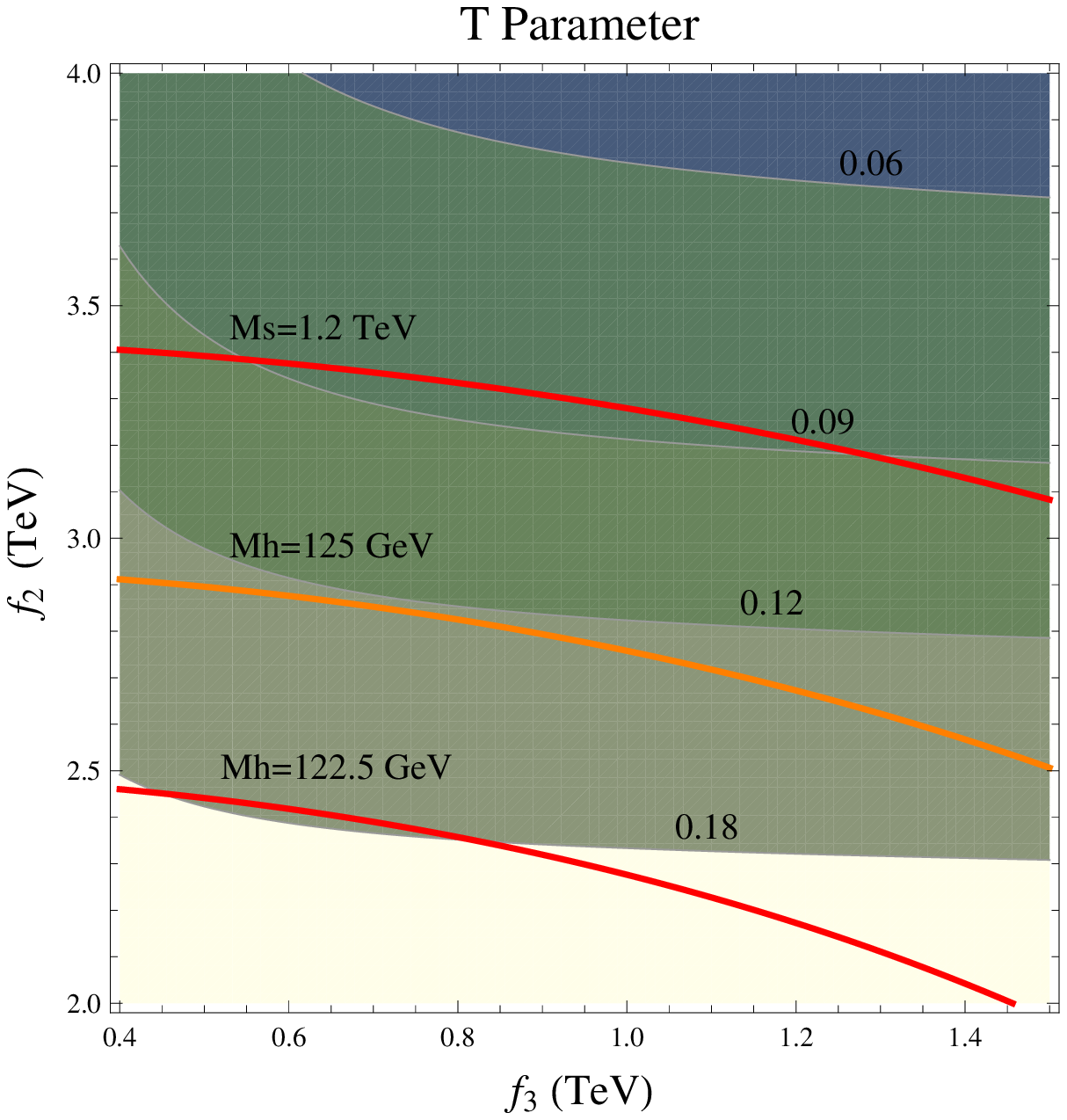}
\end{center}
\caption{ Contours of $T$ parameter as functions of two VEVs $f_2$ and $f_3$. Input parameters are: $ m_{\tilde t_R} = 300~\mbox{GeV}$,   $\tan \beta = 0.86 $, $\mu_2 = 5 ~\mbox{TeV}$, $\mu_3 = 2 ~\mbox{TeV} $, and  $ M_A = 400~\mbox{GeV} $.  In the left panel,  the gauge couplings are $\hat g_5 =1.4 $, $\hat  g_{1H} =0.378$, $\hat g_{1V} =1.5$, and $\hat g_2=0.73$.  In the right panel, the gauge couplings are $ \hat g_{5} =1.4 $, $\hat g_{1H}=0.37$, $ \hat g_{1V} =2.5$, and $\hat g_2=0.73$. The bottom red lines in both plots put a lower bound for the Higgs bosons mass under the requirement of $T < 0.18$.  The SUSY breaking scale $M_S = \sqrt {m_{\vec Q} m_{\tilde t_R}  }$ is set to be $1.2$ TeV, corresponding to the top red lines in both plots,  which puts an upper bound for the parameter space. \label{LightT}}
\end{figure}
The presence of a heavy Higgs boson with  a mass much larger than $120
~\mbox{GeV}$ will give a negative contribution to $T$ parameter
which may lead to a confliction with the experimental constraints.
It is good for us that the mixing of gauge bosons instead drives
the $T$ parameter in the positive direction so that two effects may
balance  with each other and we can go back into the consistent
region in the  $S-T$ plane.  In Fig. [\ref{HTPara}],  the
contribution to  $T$ parameter combining  both Eq.~[\ref{TG}] and
Eq.~[\ref{TH}]  is plotted against the VEVs of $f_2$ and $f_3$.
The lowest contour corresponds to the positive $T =0.18 $ bound.
The requirement of  $T < 0.18$ gives a lower bound to both the
mass of vector top partner and the mass of the Higgs bosons. Since
we demand less percentage of  fine tuning,   the SUSY breaking
scale defined as $M_S = \sqrt {m_{\vec Q} m_{\tilde t_R} }$ is
required to be around $\mathcal{O}$(1) TeV.  This latter requirement fixes the
upper bound for Higgs bosons in this model.  In the left panel of
Fig.~[\ref{HTPara}], the gauge coupling constants are taken to be
$ \hat g_5 =1.2$, $ \hat g_{1H} =0.378$, $\hat g_{1V} =1.5$, and
$\hat g_2=0.78$. The mass of right handed stop is  $300 ~
\mbox{GeV}$, $\mu$-term masses are $\mu_2 = 5 ~\mbox{TeV}$,
$\mu_3 = 2 ~\mbox{TeV}$ and the input mass parameter for Higgs
bosons in Eq.  [\ref{Higgs}]  is  $ M_A =800~\mbox{GeV}$.  The
$\tan \beta$ is taken to be $2.0$. This set of chosen parameters
predicts  the Higgs mass to be in the range of  $(188.5~\mbox{GeV} ,
~ 194 ~\mbox{GeV})$,  related to a vector top partner with its
mass in the range of $( 2.6~ \mbox{TeV} , ~ 4.8 ~\mbox{TeV}) $.
In the right panel of Fig. [\ref{HTPara}], we take another set of
gauge coupling constants $\hat g_5 =1.2$, $ \hat g_{1H} =0.37$,
$\hat g_{1V} =2.5$, and $\hat g_2=0.78$,  with  the  other input
parameters unchanged.  As we can see,  increasing $\hat g_{1V}$
gauge coupling will reduce the $T$ parameter's dependence on the
value of $ f_3$ parameter, i.e. the contour becomes more flat in
the right panel,  which will relax the lower bound of the vector
top partner  to be $m_{\vec{Q} } > 2.55 ~\mbox{TeV} $ and
therefore result in less radiative correction to the Higgs mass.
However increasing $\hat g_{1V} $ gauge coupling at the same time
enlarges the tree level Higgs quartic coupling through the
$\Delta_Y$ parameter so that the total effect is that with a
larger $\hat g_{1V} $ coupling  the Higgs bosons mass is increased
by just $1-2 ~\mbox{GeV} $ to be located in a range of $(190
~\mbox{GeV}, ~195 ~\mbox{GeV})$.  Varying the mass of the
right handed stop by hundred GeVs  can result in the lower bound
of  the Higgs mass slightly changing by a few GeVs.  As shown in
Table ~\ref{higgs}, for a specific set of the gauge couplings:
$\hat g_5 =1.2$, $\hat g_{1H} =0.378$, $\hat g_{1V} =1.5$, and
$\hat g_2=0.78$, when the right handed stop mass is varied from
$300~\mbox{GeV }$ to $550~\mbox{ GeV} $,  the  lower bound for
the Higgs bosons mass which satisfies the requirement of  $T< 0.18$  will change accordingly from $ 188.5~\mbox{GeV} $ to $ 200~\mbox{GeV}$.  The parameter space  constrained by the oblique parameter does not  exclude  a light Higgs boson, which is
preferred by current ATLAS and CMS search results.  Recent LHC
experiments observed an excess of events at $125~\mbox{GeV}$ in
the final state of $\gamma \gamma$ hence give a hint that a
Standard Model like Higgs may exist in the mass window of
$123~\mbox{GeV}- 127~\mbox{GeV}$.  The light Higgs scenario can be
achieved by tuning the $\tan \beta $. When we take $\tan \beta =
0.86$,  the Higgs mass  is limited to be $m_h \ge 122.5~\mbox{GeV} $
depending on the specific  gauge couplings and other input
parameters as shown in Fig. [\ref{LightT}].  But it will require a
heavier vector top partner with its mass larger than $3.5 ~
\mbox{TeV}$ to be consistent with the $T$ parameter constraints.

Another important constraint comes from the corrections to the $Z \to b \bar b$ vertex. The $b_L$ quark in this model is a linear combination of several fields and is mostly the gaugino of $SU(5)$, while the right handed bottom quark residing in the first moose site is only gauged with $SU(3) \times SU(2) \times U(1) $.  The left handed bottom quark couples to the heavy gauge bosons in a way different from the right handed bottom quark.  The gauge bosons mixing through Higgs VEVs induces both a correction to $Z b_L \bar b_L$ coupling  and a correction to $Z b_R \bar b_R$ coupling:
\begin{figure}[htb]
\begin{center}
\includegraphics[width=0.48\hsize]{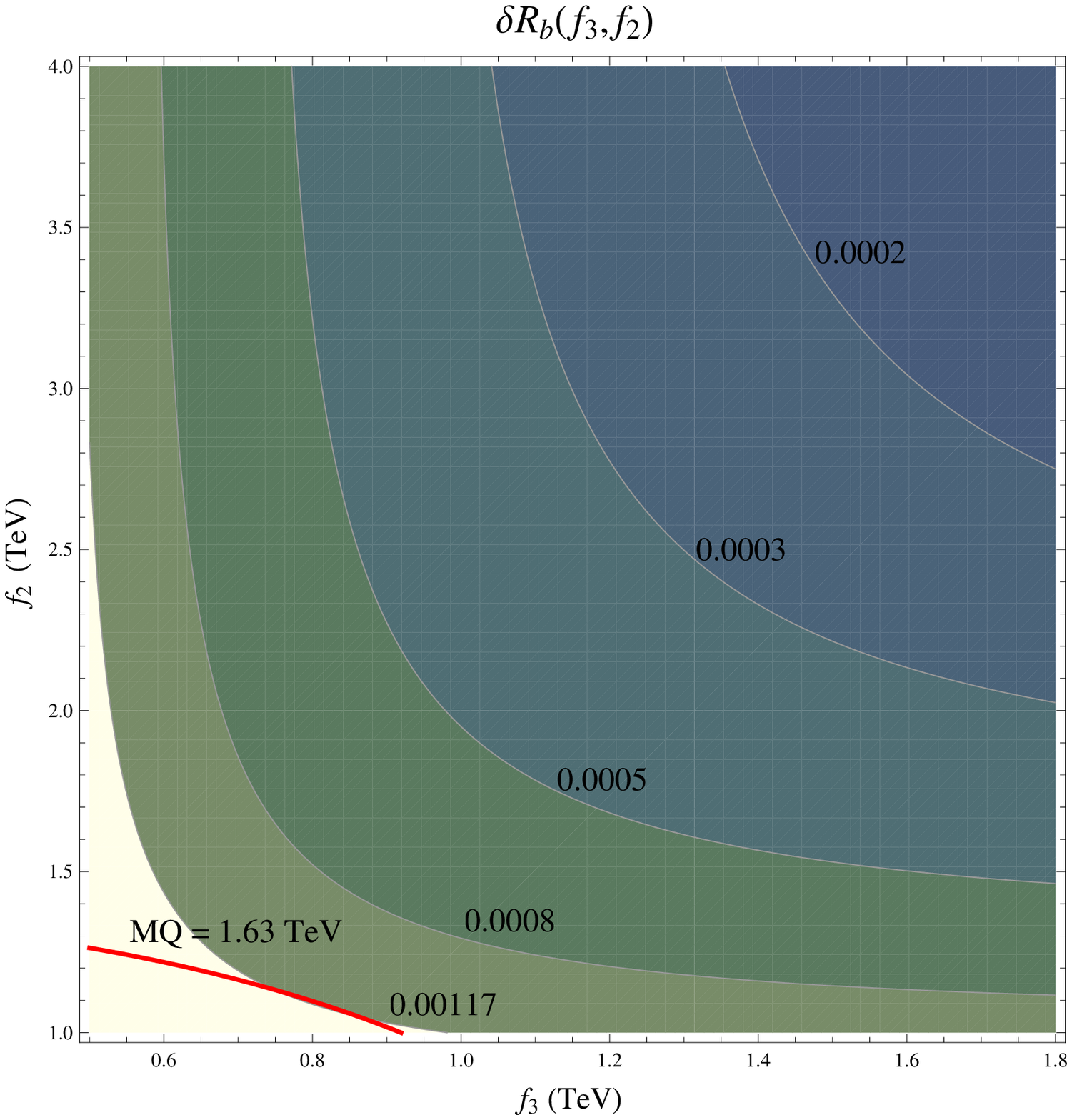}
\includegraphics[width=0.48\hsize]{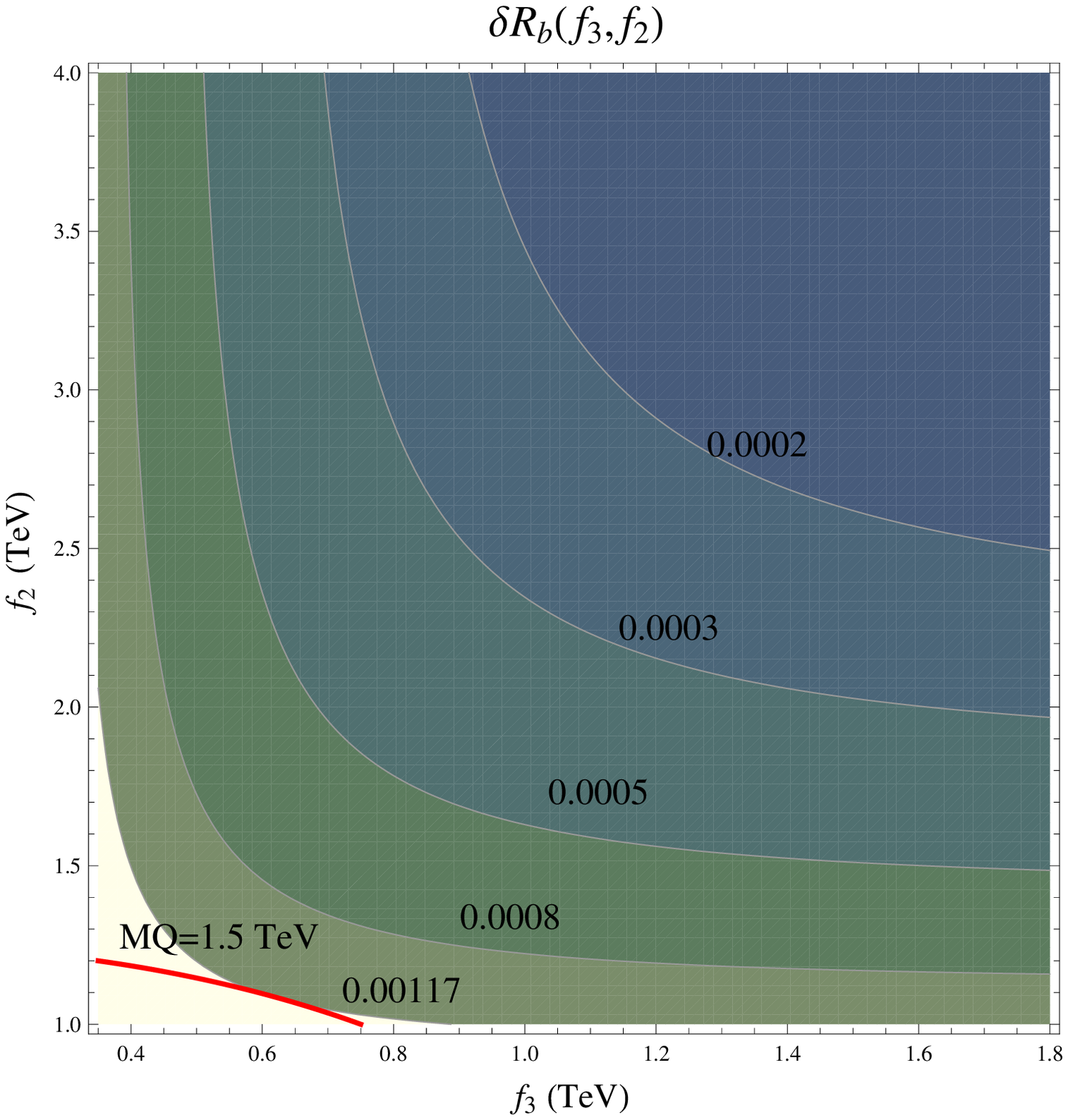}
\end{center}
\caption{ contours of $R_b$ as functions of  two VEVs $f_2$ and $f_3$. In the left panel, the gauge couplings are $ \hat g_{5} =1.2$, $\hat g_{1H}=0.378$,   $ \hat g_{1V} =1.5$, and $\hat g_{2} = 0.78 $. in the right panel the gauge couplings are  $ \hat g_{5} =1.2$, $\hat g_{1H}=0.37$, $ \hat g_{1V} =2.5$  and $\hat g_{2}= 0.78 $. The red lines give the lower bound for the mass of the vector top partner under the  requirement of  $R_b < 0.00117 $. \label{Rb}}
\end{figure}
\beq
\delta g_{z{b_L}{b_L}} = & -& \frac{e}{sc} z_\mu J_3^\mu \left( \frac{ \hat g_5^4}{ (\hat g_2^2 +\hat g_5^2)^2}\frac{v^2}{8 f_2^2 } \right)
 - \frac{e}{sc} z_\mu J_Y^\mu \left( \frac{45 \hat g_5^2 \hat g_{1H}^2 \hat g_{1V}^2\left(15 \hat g_5^2 +\hat  g_{1V}^2 \right)}{\left( 15 \hat g_5^2 \hat g_{1V}^2 + \hat g_{1H}^2 \left(15 \hat g_5^2 +\hat  g_{1V}^2 \right) \right)^2}\frac{v^2}{4 f_3^2 } \right)  \nonumber \\  \nonumber \\
 &-& \frac{e}{sc} z_\mu J_Y^\mu \left(\frac{15 \hat g_5^2\left(5 \hat g_{1H}^2 + 3 \hat g_{1V}^2 \right)\left( 15 \hat g_5^2 \hat g_V^2 + \hat g_{1H}^2\left( - 15 \hat g_5^2 + 4 \hat g_{1V}^2 \right) \right)}{\left( 15 \hat g_5^2 \hat g_{1V}^2 + \hat g_{1H}^2 \left(15 \hat g_5^2 + \hat g_{1V}^2 \right) \right)^2}\frac{v^2}{ 8 f_2^2 } \right) \label{ZBL} ~,
\eeq
\beq
\delta {g_{z{b_R}{b_R}}} &=&  - \frac{e}{{sc}}{z_\mu }J_Y^\mu \left( {\frac{{3\hat g_{1H}^2\left( {10\hat g_5^2 - \hat g_{1V}^2} \right)\left( {15\hat g_5^2\hat g_{1V}^2 +\hat g_{1H}^2 \left( { - 15\hat g_5^2 + 4\hat g_{1V}^2} \right)} \right)}}{{\left( {15\hat g_5^2\hat g_{1V}^2 + \hat g_{1H}^2\left( {15\hat g_5^2 + \hat g_{1V}^2} \right)} \right)}^2}} \frac{v^2}{8 f_2^2}\right) \nonumber \\
 && + \frac{e}{{sc}}{z_\mu }J_Y^\mu \left( {\frac{{3\hat g_{1H}^4{{\left( {15\hat g_5^2 + \hat g_{1V}^2} \right)}^2}}}{{\left( {15\hat g_5^2\hat g_{1V}^2 + \hat g_{1H}^2\left( {15\hat g_5^2 + \hat g_{1V}^2} \right)} \right)}^2}} \frac{v^2}{4f_3^2}\right) \label{ZBR} ~ .
\eeq
The expression shows that the  correction to $Z$ gauge bosons coupled to the right handed bottom quark  is much less  as it is proportional to $\hat g_{1H}^2$,  and the value of  $\hat g_{1H} $  is assumed to be small in this model .  Constraint from  $Z \to b \bar b $  is measured by the  branch ratio of  $R_b = \Gamma(Z \to b \bar b ) / \Gamma(\mbox{hadron})$.  The deviation of $R_b$ due to the new physics can be expressed in terms of $\delta g_L^{NP} $ and $\delta g_R^{NP}$ :

\beq
\delta {R_b} &=& 2{R_b}(1 - {R_b}) \left (\frac{{{g_L}}}{{g_L^2 + g_R^2}}\delta g_L^{NP} + \frac{{{g_R}}}{{g_L^2 + g_R^2}}\delta g_R^{NP}\right)   \label{Zb} \\
 g_L &=& -\frac{1}{2} + \frac{1}{3} s^2    \qquad  \qquad  g_R = \frac{1}{3} s^2
\eeq
here $R_b$ is the SM value  predicted by the electroweak fit and
its value is $R_b = 0.21578 + 0.0005 (-0.0008)$.  The deviation $
\delta R_b$ ,  used to describe  the  difference between its
observation value and the SM fit result,  is given by  the
experimental measurement  \cite{EW}:
\beq
\delta {R_b} &=& 0.00051 \pm 0.00066  \eeq
Substituting Eq.~[\ref{ZBL}] and Eq.~[\ref{ZBR}] into
Eq.~[\ref{Zb}], and we plot the dependence of $\delta {R_b} $ on
the two VEV parameters $f_2$ and $f_3$ in Fig. [\ref{Rb}].  The
lowest contour in that figure corresponds to the upper bound of
$\delta R_b = 0.00117$  ,  which gives a loose bound  on the mass
of the  vector top partner compared with the T parameter
constraint.  For comparison reason, we will adopt the same two
sets of gauge couplings to evaluate the value of $\delta{R_b}$ ,
as we do in analyzing the T parameter.  In the left panel of Fig.
[\ref{Rb}] ,  the gauge couplings are :  $ \hat g_{5} =1.2$,
$\hat g_{1H}=0.378$,  $ \hat g_{1V} =1.5$, and $ \hat g_2 = 0.78$, 
by requiring $\delta R_b < 0.00117$,  the mass of the vector
top partner  is limited to be  $m_{\vec Q} \ge  1.63 ~
\mbox{TeV}$.  In the right panel,  gauge couplings are taken to be
$ \hat g_{5} =1.2$, $\hat g_{1H}=0.37$, $ \hat g_{1V} =2.5$ and
$\hat g_2 =0.78 $.  Since both $ \delta {g_{z{b_L}{b_L}}}  $ and
$\delta {g_{z{b_R}{b_R}}}  $ will decrease as we increase the
value of $\hat  g_{1V}$,  we  could have a smaller bound value
$m_{\vec Q} \ge  1.5 ~ \mbox{TeV}$  for the vector top partner.

It can be seen  that  the $T$ parameter measuring the amount of custodial symmetry breaking  constrains the parameter space in a more stringent way.  By contrast the measurement from $Z \to b \bar b $ gives a rather loose and negligible bound for the mass of the vector top partner in this model.  Let us assume that $\hat g_5$, $\hat g_{1V} $  are relatively big, and  $\hat g_{1H}$ is much smaller,  through tuning the other parameters,  the theory is capable to accommodate  a  Higgs bosons with its mass in the range of  $( 122.5 ~\mbox{GeV}, ~200 ~\mbox{GeV} )$, after considering the electroweak constraints.  Notice that in  the case  of   a  light Higgs boson,  we generally demand $\tan \beta \sim (0.8-0.9)$  and  a  large $m_{\vec Q}$ in order to satisfy  the $T< 0.18$ requirement.

\section{ Conclusions }
In this paper, I present that adding extra singlet chiral
superfields to interact with the link fields can trigger the gauge
symmetry breaking and with an appropriately  arranged
superpotential,  the VEVs will  lead to spontaneous Supersymmetry
breaking at the same time.  I also add two chiral superfields
transforming under the $SU(2) $ adjoint representation and the $
SU(3) $ adjoint representation respectively to lift the moduli so
that there is no light mode after the gauge symmetry breaking. Due
to the nondecoupling D-term effects of those heavy fields,  a
larger Higgs quartic coupling is obtained. We explicitly
demonstrate that in the large SUSY breaking limit, the effective
low energy D-term is the same as in the unbroken gauge theory.
Since  the gauge couplings of  the extra gauge groups $SU(5)
\times U(1)_{1V} $ under which the Higgs boson are charged are
taken to be strong, with a moderate $O(1) $  $\tan \beta $  the
Higgs mass is heavy enough at the tree level. After taking the
radiative corrections into account, the Higgs mass can be raised
to be well beyond the LEP bound.

\newpage

\acknowledgments
I thank Qing-Hong Cao for discussion of this model and reading this manuscript.  This work is in part supported by the National Natural Science Foundation of China (Grants No. 11021092, No.11075002, and No.11075011) .

\end{document}